\documentclass[aps,twocolumn,showpacs,preprintnumbers,amsmath,amssymb]{revtex4}
\usepackage{xspace}

\usepackage{graphicx}% Include figure files
\usepackage{bm}% bold math

\newcommand{\etal}{\textit{et al.}\xspace}
\newcommand{\tc}{\ensuremath{T_{\rm{c}}}}            % In a sentence use $\tc$ for space ok
\newcommand{\hc}{\ensuremath{H_{\rm{c2}}}}           % In a sentence use $\hc$ for space ok
\newcommand{\mhc}{\ensuremath{\mu_0H_{\rm{c2}}}}     % In a sentence use $\mhc$ for space ok
\newcommand{\rhn}{\ensuremath{\rho_{\rm{n}}}}        % In a sentence use $\rhn$ for space ok
\newcommand{\jc}{\ensuremath{J_{\rm{c}}}}            % In a sentence use $\jc$ for space ok
\newcommand{\nef}{\ensuremath{N(E_{\rm{F}})}}        % In a sentence use $\nef$ for space ok

\begin{document}

\title[A Review of the Properties of Nb$_3$Sn.]{A Review of the Properties of Nb$_3$Sn and Their Variation with A15 Composition, Morphology and Strain State}
\author{A. Godeke\footnote{Previous address: Low Temperature Division, Faculty of Science and Technology, University of Twente, P.O. Box 217, 7500~AE, Enschede, The Netherlands}}
\affiliation{Ernest Orlando Lawrence Berkeley National Laboratory, Berkeley, CA 94720}
%\ead{agodeke@lbl.gov} %IOP

\date{\today }

%------------------------------------------------------------------------------------------------------------

\begin{abstract}
Significant efforts can be found throughout the literature to optimize the current carrying capacity of Nb$_3$Sn superconducting wires. The achievable transport current density in wires depends on the A15 composition, morphology and strain state. The A15 sections in wires contain, due to compositional inhomogeneities resulting from solid state diffusion A15 formation reactions, a distribution of superconducting properties. The A15 grain size can be different from wire to wire and is also not necessarily homogeneous across the A15 regions. Strain is always present in composite wires, and the strain state changes as a result of thermal contraction differences and Lorentz forces in magnet systems. To optimize the transport properties it is thus required to identify how composition, grain size and strain state influence the superconducting properties. This is not accurately possible in inhomogeneous and spatially complex systems such as wires. This article therefore gives an overview of the available literature on simplified, well defined (quasi-)homogeneous laboratory samples. After more than 50 years of research on superconductivity in Nb$_3$Sn, a significant amount of results are available, but these are scattered over a multitude of publications. Two reviews exist on the basic properties of A15 materials in general, but no specific review for Nb$_3$Sn is available. This article is intended to provide such an overview. It starts with a basic description of the Niobium-Tin intermetallic. After this it maps the influence of Sn content on the the electron-phonon interaction strength and on the field-temperature phase boundary. The literature on the influence of Cu, Ti and Ta additions will then be briefly summarized. This is followed by a review on the effects of grain size and strain. The article is concluded with a summary of the main results.
\end{abstract}

\pacs{74.25.Dw, 74.62.Bf, 74.62.Dh, 74.81.Bd\\ \\Accepted Invited Topical Review for Superconductor Science and Technology\\ http://www.iop.org/EJ/journal/SUST \copyright IOPP 2006\\Provisionally scheduled for July 2006
}

\maketitle
%
%-------------------------------------------------------------------------------------------------------------
\section{\label{Introduction}Introduction}
Superconductivity in Nb$_3$Sn was discovered by Matthias \etal in 1954~\cite{Matthias1954PR}, one year after the discovery of V$_3$Si, the first superconductor with the A15 structure by Hardy and Hulm in 1953~\cite{Hardy1953PR}. Its highest reported critical temperature is 18.3~K by Hanak \etal in 1964~\cite{Hanak1964RCA}. Ever since its discovery, the material has received substantial attention due to its possibility to carry very large current densities far beyond the limits of the commonly used NbTi. It has regained interest over the past decade due to the general recognition that NbTi, the communities' present workhorse for large scale applications, is operating close to its intrinsic limits and thus exhausted for future application upgrades. Nb$_3$Sn approximately doubles the available field-temperature regime with respect to NbTi and is the only superconducting alternative that can be considered sufficiently developed for large scale applications.

\begin{figure*}
\includegraphics [scale=1]{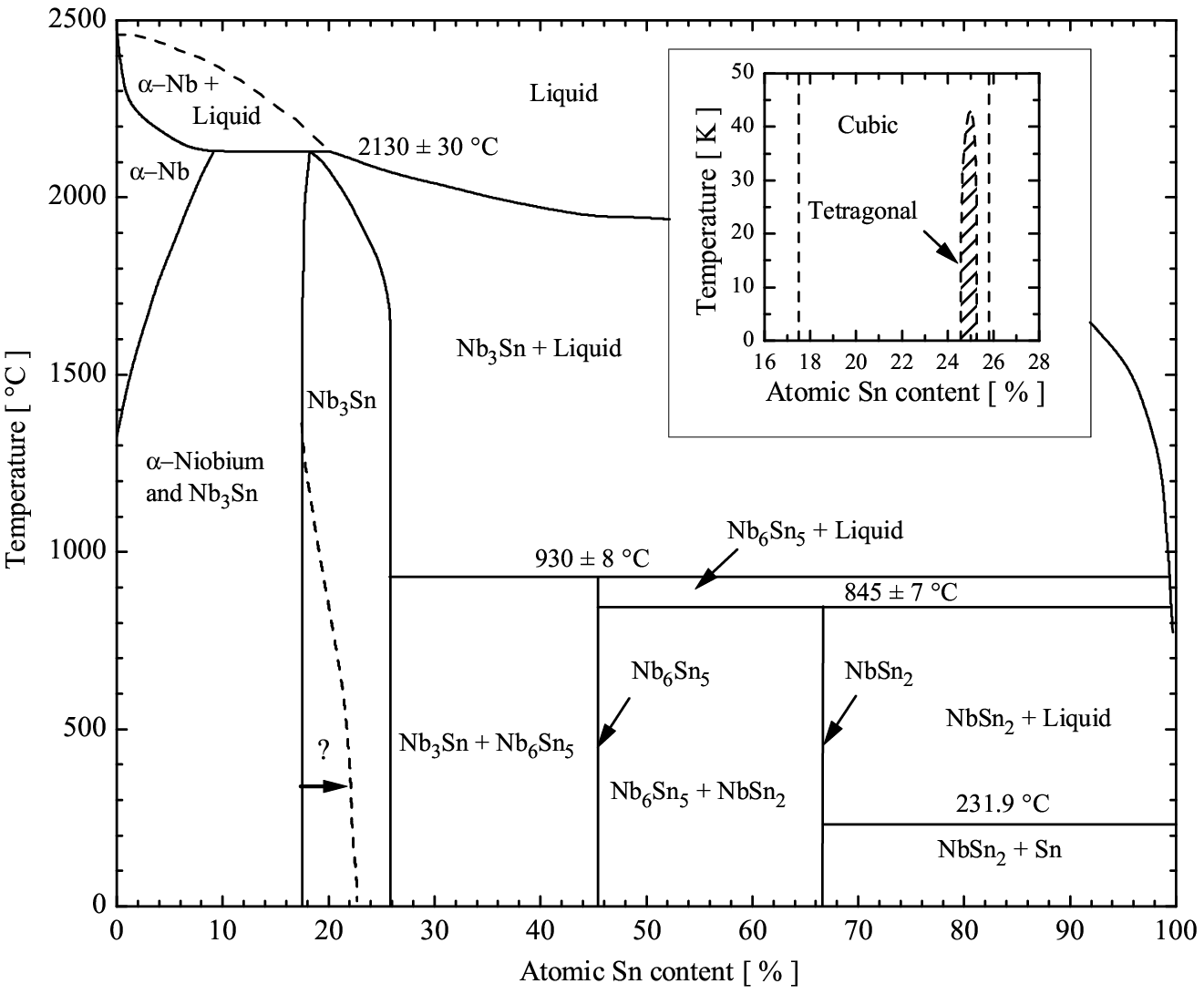}
\caption{\label{Charlesworth} Binary phase diagram of the Nb-Sn system after Charlesworth \etal~\cite{Charlesworth1970JMS}$^{\dag}$, with an optional modification to suggested preference for Sn rich A15 formation~\cite{Devantay1981JMS,Jewell2004ACEM,Vieland1964RCA,Toffolon1998JPE,Toffolon2002JPE,Schiffman1982HTS}. The inset depicts the low temperature phase diagram after Fl\"{u}kiger~\cite{Flukiger1982ACEM}$^\ddag$, which includes the stability range of the tetragonal phase.\\$^{\dag}$\scriptsize{\textit{\copyright 1970 Chapman and Hall Ltd. Adapted with kind permission of Springer Science and Business Media. There are instances where we have been unable to trace or contact the original authors. If notified the publisher will be pleased to rectify any errors or omissions at the earliest opportunity.}}\\$^{\ddag}$\scriptsize{\textit{\copyright 1982 Plenum Press. Adapted with kind permission of Springer Science and Business Media and R.~Fl\"{u}kiger.}}}
\end{figure*}

Intermetallic Niobium-Tin is based on the superconductor Nb, which exists in a \textit{bcc} Nb structure ($\tc \cong9.2~\rm{K}$), or a metastable Nb$_3$Nb A15 structure ($\tc\cong5.2~\rm{K}$)~\cite{Flukiger1987TECH,Stewart1980PRB}. When alloyed with Sn and in thermodynamic equilibrium, it can form either Nb$_{1-\beta}$Sn$_\beta$  (about $0.18 \leq \beta \leq 0.25$) or the line compounds Nb$_6$Sn$_5$ and NbSn$_2$ according to the generally accepted binary phase diagram by Charlesworth \etal~\cite{Charlesworth1970JMS} (figure~\ref{Charlesworth}). The solid solution of Sn in Nb at low concentrations ($\beta < 0.05$) gradually reduces the critical temperature of \textit{bcc} Nb from about 9.2~K to about 4~K at $\beta = 0.05$ (Fl\"{u}kiger in \cite{Fonerbook1981}). Both the line compounds at $\beta = 0.45$ and 0.67 are superconducting with $\tc < 2.8~\rm{K}$ for Nb$_6$Sn$_5$~\cite{Enstrom1966JAP,Charlesworth1966PL} and $\tc < 2.68~\rm{K}$ for NbSn$_2$~\cite{Charlesworth1966PL,Ooijen1962PL} and thus are of negligible interest for practical applications.

The Nb-Sn phase of interest occurs from $\beta \cong 0.18$ to 0.25. It can be formed either above 930~$^\circ\rm{C}$ in the presence of a Sn-Nb melt, or below this temperature by solid state reactions between Nb and Nb$_6$Sn$_5$ or NbSn$_2$. Some investigations suggest that the nucleation of higher Sn intermetallics is energetically more favorable for lower formation temperatures~\cite{Devantay1981JMS,Jewell2004ACEM,Vieland1964RCA,Toffolon1998JPE,Toffolon2002JPE,Schiffman1982HTS}. This is indicated by the dashed line within the Nb$_{1-\beta}$Sn$_\beta$ stability range in figure~\ref{Charlesworth}. The critical temperature for this phase depends on composition and ranges approximately from 6 to 18~K~\cite{Devantay1981JMS}. At low temperatures and at $0.245 < \beta < 0.252$ it can undergo a shear transformation at $T_{\rm{m}} \cong 43~\rm{K}$. This results in a tetragonal structure in which the reported ratio of the lattice parameters $c / a = 1.0026$ to 1.0042~\cite{Flukiger1982ACEM,DewHughes1975CRY,Mailfert1967PLA,Mailfert1969PSS,Chu1974JLTP,Vieland1970JPCS,Junod1978JPCS,Vieland1971PLA}. This transformation is schematically depicted in the low temperature extension of the binary phase diagram as shown in the inset in figure~\ref{Charlesworth}.
\begin{figure}
\includegraphics [scale=0.8]{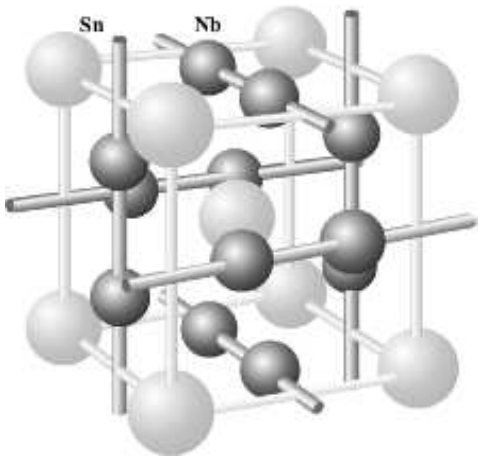}
\caption{\label{Unitcell} Schematic presentation of the Nb$_3$Sn A15 unit cell. The light spheres represent Sn atoms in a \textit{bcc} lattice. The dark spheres represent orthogonal chains of Nb atoms bisecting the \textit{bcc} cube faces.}
\end{figure}

Nb$_{1-\beta}$Sn$_\beta$ exists in the brittle A15 crystal structure with a cubic unit cell, as schematically depicted in figure~\ref{Unitcell}. The Sn atoms form a \textit{bcc} lattice and each cube face is bisected by orthogonal Nb chains. The importance of these chains is often emphasized to qualitatively understand the generally high critical temperatures of A15 compounds. An excellent review on this subject is given by Dew-Hughes in~\cite{DewHughes1975CRY} and can be summarized specifically for the Nb-Sn system. In \textit{bcc} Nb the shortest spacing between the atoms is about 0.286~nm, starting from a lattice parameter $a=0.330~\rm{nm}$~\cite{Straumanis1970JAC}. In the A15 lattice, with a lattice parameter of about 0.529~nm for the stoichiometric composition~\cite{Devantay1981JMS}, the distance between the Nb atoms is about 0.265~nm. This reduced Nb distance in the chains is suggested to result in a narrow peak in the d-band density of states (DOS) resulting in a very high DOS near the Fermi level. This is in turn believed to be responsible for the high $\tc$ in comparison to \textit{bcc} Nb. Variations in $\tc$ are often discussed in terms of long-range crystallographic ordering~\cite{Flukiger1987TECH,DewHughes1975CRY,Flukiger1984ACEM,Flukiger1987TM,Beasley1982ACE} since deviations in the Nb chains will affect the DOS peak. Tin deficiency in the A15 structure causes Sn vacancies, but these are believed to be thermodynamically unstable~\cite{Flukiger1987TECH,Welch1984JPCS}. The excess Nb atoms will therefore occupy Sn sites, as will be the case with anti-site disorder. This affects the continuity of the Nb chains which causes a rounding-off of the DOS peak. Additionally, the \textit{bcc} sited Nb atoms cause there own broader d-band, at the cost of electrons from the Nb chain peak.

The above model is one of many factors that are suggested to explain effects of disorder on the superconducting transition temperature~\cite{Beasley1982ACE}. It is presented here since it intuitively explains the effects of Nb chain atomic spacing and deviations from long range crystallographic ordering and also is mostly cited. In the Nb$_3$Nb system with a lattice constant of 0.5246~nm~\cite{Stewart1980PRB} the distance between the chained Nb atoms of about 0.262~nm is, as in the A15 Nb-Sn phase, lower than in the \textit{bcc} lattice. This would suggest an increased $\tc$ compared to \textit{bcc} Nb. Its lower $\tc$ of 5.2~K~\cite{Flukiger1987TECH} could possibly be explained in the above model by assuming that 'Sn sited' Nb atoms create their own d-band at the cost of electrons from the chained Nb atoms, thereby reducing their DOS peak and thus degrading the expected $\tc$ gain.

%-------------------------------------------------------------------------------------------------------------

\section{\label{Variationsinlatticeproperties}Variations in lattice properties}
Variations in superconducting properties of the A15 phase are throughout the literature related to variations in the lattice properties through the lattice parameter ($a$), atomic Sn content ($\beta$), the normal state resistivity just above $\tc$ ($\rhn$) or the long range order (LRO). The latter can be defined quantitatively in terms of the Bragg-Williams order parameters $S_{\rm{a}}$ and $S_{\rm{b}}$ for the chain sites and the cubic sites respectively. These can be expressed in terms of occupation factors (\cite{Flukiger1987TECH}, Fl\"{u}kiger in \cite{Fonerbook1981}):
\begin{equation}\label{orderparameters}
    S_{\rm{a}}  = \frac{{r_{\rm{a}}  - \beta }}
    {{1 - \beta }}{\rm{~~and~~}}S_{\rm{b}}  = \frac{{r_{\rm{b}}  - \left( {1 - \beta } \right)}}
    {{1 - \left( {1 - \beta } \right)}},
\end{equation}
where $r_{\rm{a}}$ and $r_{\rm{b}}$ are the occupation factors for A atoms at the chain sites and B atoms at the cubic sites respectively in an A15 $\rm{A}_{1-\beta}\rm{B}_\beta$ system. At the stoichiometric composition $S=S_{\rm{a}}=S_{\rm{b}}=1$ and $S=0$ represents complete disorder. Qualitatively, the more general term 'disorder' is used for the LRO since any type of disorder (e.g. quenched in thermal disorder, off-stoichiometry, neutron irradiation) reduces $\tc$~\cite{Beasley1982ACE}. A review of the literature suggests that the aforementioned lattice variables ($a$, $\beta$ and $S$) and $\rhn$ are, at least qualitatively, interlinked throughout the A15 phase composition range. This makes it possible to relate them back to the main available parameter in multifilamentary composite wires, i.e. the atomic Sn content.
\begin{figure}
\includegraphics [scale=1]{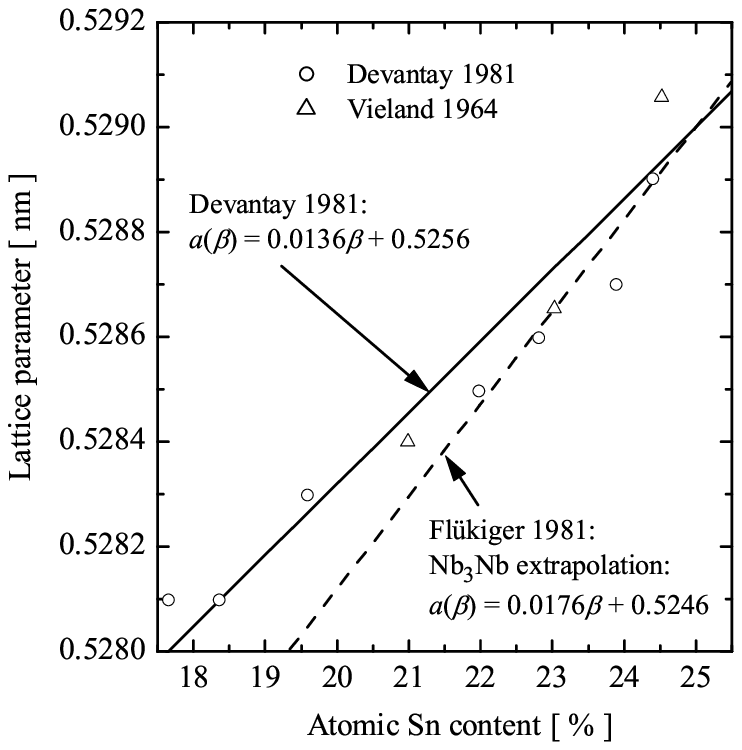}
\caption{\label{latticeparametersSncontent} Lattice parameter as function of atomic Sn content after Devantay \etal~\cite{Devantay1981JMS}$^{\dag}$, including proposed linear dependencies after Devantay \etal and Fl\"{u}kiger in~\cite{Fonerbook1981}$^{\ddag}$.\\$^{\dag}$\scriptsize{\textit{\copyright 1981 Chapman and Hall Ltd. Adapted with kind permission of Springer Science and Business Media and R.~Fl\"{u}kiger.}}\\$^{\ddag}$\scriptsize{\textit{\copyright 1981 Plenum Press. Adapted with kind permission of Springer Science and Business Media and R.~Fl\"{u}kiger.}}}
\end{figure}
\begin{figure}
\includegraphics [scale=1]{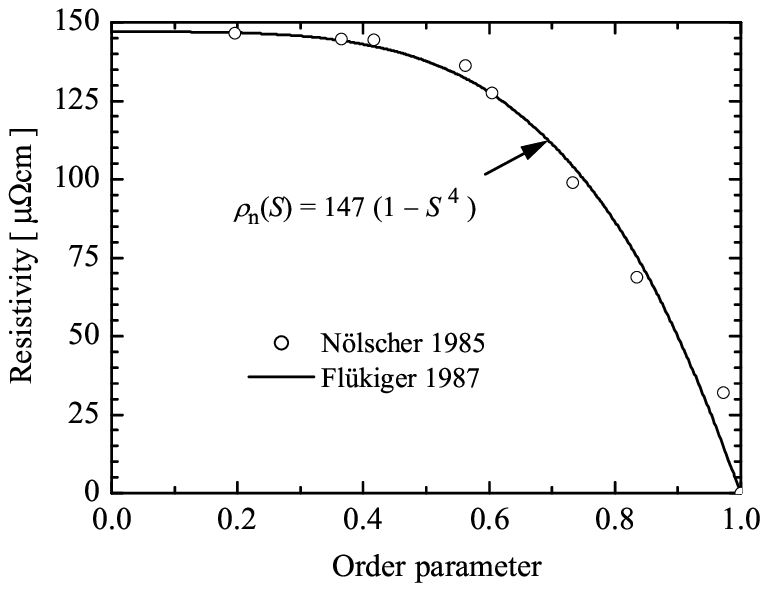}
\caption{\label{rhoorder} Resistivity as function of order parameter including a fourth power fit after Fl\"{u}kiger \etal~\cite{Flukiger1987TM}$^{\dag}$.\\$^{\dag}$\scriptsize{\textit{\copyright 1987 IEEE. Adapted with kind permission of IEEE and R.~Fl\"{u}kiger.}}}
\end{figure}

The lattice parameter as function of composition was measured by Devantay \etal~\cite{Devantay1981JMS} and combined with earlier data from Vieland~\cite{Vieland1964RCA}. These data are reproduced in figure~\ref{latticeparametersSncontent}. The solid line is a fit to the data similar as given by Devantay \etal:
\begin{equation}\label{devantaylin}
    a\left( \beta  \right) = 0.0136 \beta + 0.5256 ~~ \left[ {{\rm{nm}}} \right].
\end{equation}
The dotted line is an alternative line fit proposed by Fl\"{u}kiger in \cite{Fonerbook1981}:
\begin{equation}\label{flukigerlin}
    a\left( \beta  \right) = 0.0176 \beta + 0.5246 ~~ \left[ {{\rm{nm}}} \right],
\end{equation}
in which the lattice parameter is extrapolated from 0.529~nm at $\beta=0.25$ to a Nb$_3$Nb value of 0.5246~nm~\cite{Stewart1980PRB} at $\beta=0$. The argument for doing so is that from analysis of the lattice parameter of a wide range of Nb$_{1-\beta} \rm{B}_\beta$ superconductors as function of the amount of B element, it appears that most lattice parameters extrapolate to the Nb$_3$Nb value for $\beta=0$. It is interesting to note that the rise in lattice parameter with Sn content is apparently contrary to all other A15 compounds, for which a reduction in lattice parameter is observed with increasing B-element. For the analysis in this article, (\ref{devantaylin}) is presumed an accurate description for the lattice parameter results as measured by Devantay \etal and Vieland for the Nb-Sn A15 stability range.

The amount of disorder, introduced by irradiation or quenching is often related to changes in resistivity. Furthermore, the amount of disorder is important in discussions on the strain sensitivity of the superconducting properties in the Nb-Sn system in comparison to other A15 compounds (section~\ref{strain}). A relation between $\rhn$ and the order parameter $S$ in Nb$_3$Sn was established by Fl\"{u}kiger \etal~\cite{Flukiger1987TM} after analysis of ion irradiation data obtained by N\"{o}lscher and Seamann-Ischenko~\cite{Nolscher1985PRB} and data from Drost \etal~\cite{Drost1985TM}. This is reproduced in figure~\ref{rhoorder}. Fl\"{u}kiger \etal proposed a fourth power fit to describe the $\rhn(S)$ data:
\begin{equation}\label{rhos}
    \rho _{\rm{n}} \left( S \right) = 147\left( {1 - S} \right)^4 ~~ \left[ {{\rm{\mu \Omega cm}}} \right],
\end{equation}
represented by the solid curve through the data points.
\begin{figure}
\includegraphics [scale=1]{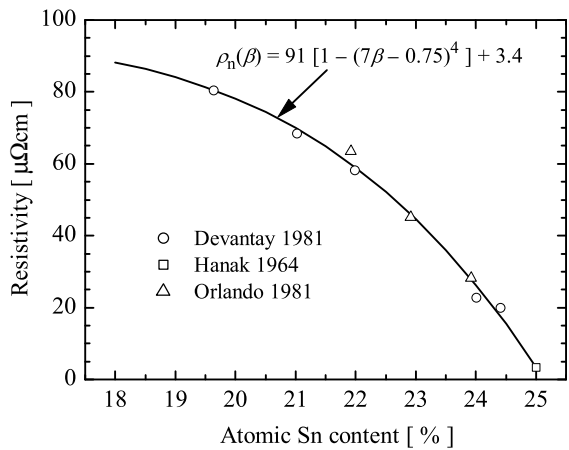}
\caption{\label{rhoSncontent} Resistivity as function of atomic Sn content after Fl\"{u}kiger~\cite{Flukiger1982BUCKELBOOK,Flukiger1987TECH}$^{\dag}$. The solid curve is a fit to the data according to (\ref{rhobeta}).\\$^{\dag}$\scriptsize{\textit{Adapted with kind permission of R.~Fl\"{u}kiger.}}}
\end{figure}

The superconducting properties of the Nb-Sn system are mostly expressed in terms of either resistivity or atomic Sn content. Resistivity data as function of composition were collected by Fl\"{u}kiger~\cite{Flukiger1987TECH,Flukiger1982BUCKELBOOK} from Devantay \etal~\cite{Devantay1981JMS}, Orlando \etal~\cite{Orlando1979PRB,Orlando1981TM}, and Hanak \etal~\cite{Hanak1964RCA} and are reproduced in figure~\ref{rhoSncontent}. The relation between the two parameters from various sources is consistent. Since Sn deficiency will result in anti-site disorder (section \ref{Introduction}) it is assumed here that $\rhn(\beta)$ will behave similarly to $\rhn(S)$, i.e. a fourth power fit, identical to (\ref{rhos}) will be appropriate. The solid curve in figure~\ref{rhoSncontent} is a fit according to:
\begin{equation}\label{rhobeta}
    \rho _{\rm{n}} \left( \beta  \right) = 91\left[ {1 - \left( {7\beta  - 0.75} \right)^4 } \right] + 3.4~~\left[ {{\rm{\mu \Omega cm}}} \right],
\end{equation}
which accurately summarizes the available data.
%
%--------------------------------------------------------------------------------------
%
\section{\label{electronphonon}Electron-phonon interaction as function of atomic tin content}
The Nb$_3$Sn intermetallic is generally referred to as a strong coupling superconductor. Any connection to microscopic formalisms, however, requires some level of understanding of the details of the interaction, since it changes the way in which physical quantities can be derived from measurements. More specifically, it requires a description for the interaction strength that varies with composition.

The BCS theory~\cite{Bardeen1957PR} provides a weak coupling approximation for the energy gap at zero temperature:
\begin{equation}\label{bcsgap}
    \Delta _0  \cong 2\hbar \omega _{\rm{c}} \exp \left[ { - \frac{1}{{\lambda _{{\rm{ep}}} }}} \right],
\end{equation}
in which $\lambda_{\rm{ep}}$ is a dimensionless electron-phonon interaction parameter~\cite{Eliashberg1960ZETF}. $\hbar \omega _{\rm{c}}$ is a cutoff energy away from the Fermi level ($E_{\rm{F}}$) outside which the attractive electron-electron interaction potential ($V_0$) becomes zero. Equation~\ref{bcsgap} is valid for $\lambda_{\rm{ep}}\ll 1$. Evaluation of the temperature dependency of the gap $\Delta(T)$ and requiring that the gap becomes zero as $T \rightarrow \tc$ yields the following description for the critical temperature in the weak coupling approximation:
\begin{equation}\label{bcstc}
    \tc \left( 0 \right) \cong \frac{{2e^{\gamma _{\rm{E}} } }}{{\pi k_{\rm{B}} }}\hbar \omega _{\rm{c}} \exp \left[ { - \frac{1}{{\lambda _{{\rm{ep}}} }}} \right],
\end{equation}
in which $\gamma _{\rm{E}}\cong0.577$ (Euler's constant). The ratio between the zero temperature gap (\ref{bcsgap}) and the zero field critical temperature (\ref{bcstc}): $2\Delta_0/k_{\rm{B}}\tc=3.528$ is a constant and represents the BCS weak coupling limit. For strong coupling (\ref{bcsgap}) and (\ref{bcstc}) are no longer valid since they become dependent on the details of the electron-phonon interaction. This effectively results in a rise of the ratio $2\Delta_0/k_{\rm{B}}\tc$ above the weak coupling limit of 3.528.
\begin{figure*}
\includegraphics [scale=1]{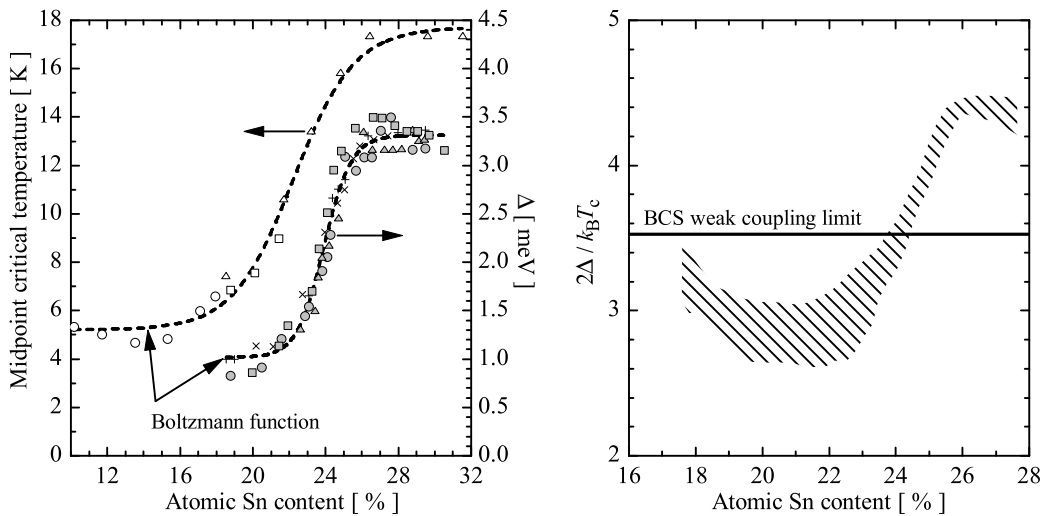}
\caption{\label{moore} Critical temperature and superconducting gap as function of composition (left plot) and the ratio $2\Delta_0/k_{\rm{B}}\tc$ as function of composition (right plot). The ratio indicates weak coupling for compositions below 23 to 24 at.\% Sn and strong coupling for compositions close to stoichiometry. After Moore \etal~\cite{Moore1979PRB}$^{\dag}$.\\$^{\dag}$\scriptsize{\textit{\copyright 1979 The American Physical Society. Adapted with kind permission of the American Physical Society and M.R.~Beasley.}}}
\end{figure*}

Moore \etal~\cite{Moore1979PRB} have analyzed the superconducting gap and critical temperature of Nb-Sn films as function of atomic Sn content by tunneling experiments. Their main result is reproduced in figure~\ref{moore}. In the left plot the inductively measured midpoint (i.e. halfway the transition) critical temperatures, and from tunneling current-voltage characteristics derived gaps are reproduced as function of composition. The dotted curves are fits to the data using a Boltzmann sigmoidal function:
\begin{equation}\label{boltzmann}
    y\left( \beta  \right) = \frac{{y_{{\rm{min}}}  - y_{{\rm{max}}} }}{{1 - \exp \left( {\frac{{\beta  - \beta _0 }}{{\rm{d}\beta }}} \right)}} + y_{{\rm{max}}},
\end{equation}
where $y$ represents $\tc$ or $\Delta$, $\beta$ is the atomic Sn content and $y_{\rm{min}}$, $y_{\rm{max}}$, $\beta_0$ and $\rm{d}\beta$ are fit parameters. Both $\tc(\beta)$ and $\Delta(\beta)$ are accurately described by (\ref{boltzmann}).

The A15 composition range as used by Moore \etal is slightly wider than generally accepted on the basis of the binary phase diagram of Charlesworth \etal~(figure~\ref{Charlesworth}, \cite{Charlesworth1970JMS}). Nevertheless, following~\cite{Moore1979PRB} the ratio $2\Delta_0/k_{\rm{B}}\tc$ can be derived from these data sets and is reproduced in the right graph in figure~\ref{moore}. The BCS weak coupling limit of 3.528 is indicated by the solid line. The weak coupling value of about 3 for low Sn content A15 was attributed by Moore \etal to the finite inhomogeneity in their samples of 1 to 1.5 at.\% Sn, in combination with an inductive measurement of $\tc$ which preferably probes the highest Sn fractions. These $\tc$ values could thus be higher than representative for the bulk. It is also known that in tunneling experiments the gap at the interface can be lower than in the bulk, effectively lowering the measured value. Also, the gap measurement was performed at finite temperature, which also reduces its value. A fourth possible origin is the existence of a second gap in Nb$_3$Sn, as was recently postulated by Guritanu \etal~\cite{Guritanu2004PRB}. Nevertheless it is clear that the Nb-Sn system shows weak coupling for most of its A15 composition range and only becomes strong coupled for compositions above 23 to 24 at.\% Sn. The general statement that Nb$_3$Sn is a strong coupling superconductor therefore only holds for compositions close to stoichiometry. Strong coupling corrections to microscopic descriptions thus become relevant for compositions approaching stoichiometry and should be accounted for in any theory that is applied to describe the entire A15 composition range.
%
%---------------------------------------------------------------------------------------------------------------
%
\section{\label{tcandhc2} $\tc$ and $\hc$ as function of atomic tin content}
Compositional gradients will inevitably occur in wires since their A15 regions are formed by a solid state diffusion process. It is therefore important to know the variation of the critical temperature and upper critical field with composition. The most complete data-sets of $\tc(\beta)$ that exist in the literature are those of Moore \etal (\cite{Moore1979PRB}, figure~\ref{moore}) and Devantay \etal~\cite{Devantay1981JMS}. The data of Moore \etal suggest, as mentioned in the previous section, a slightly broader A15 composition range than is generally accepted to be stable according to the binary phase diagram by Charlesworth \etal (\cite{Charlesworth1970JMS}, figure~\ref{Charlesworth}). The data from Devantay \etal, however, are in agreement with the accepted A15 stability range and are therefore assumed to be more accurate.
\begin{figure*}
\includegraphics [scale=1]{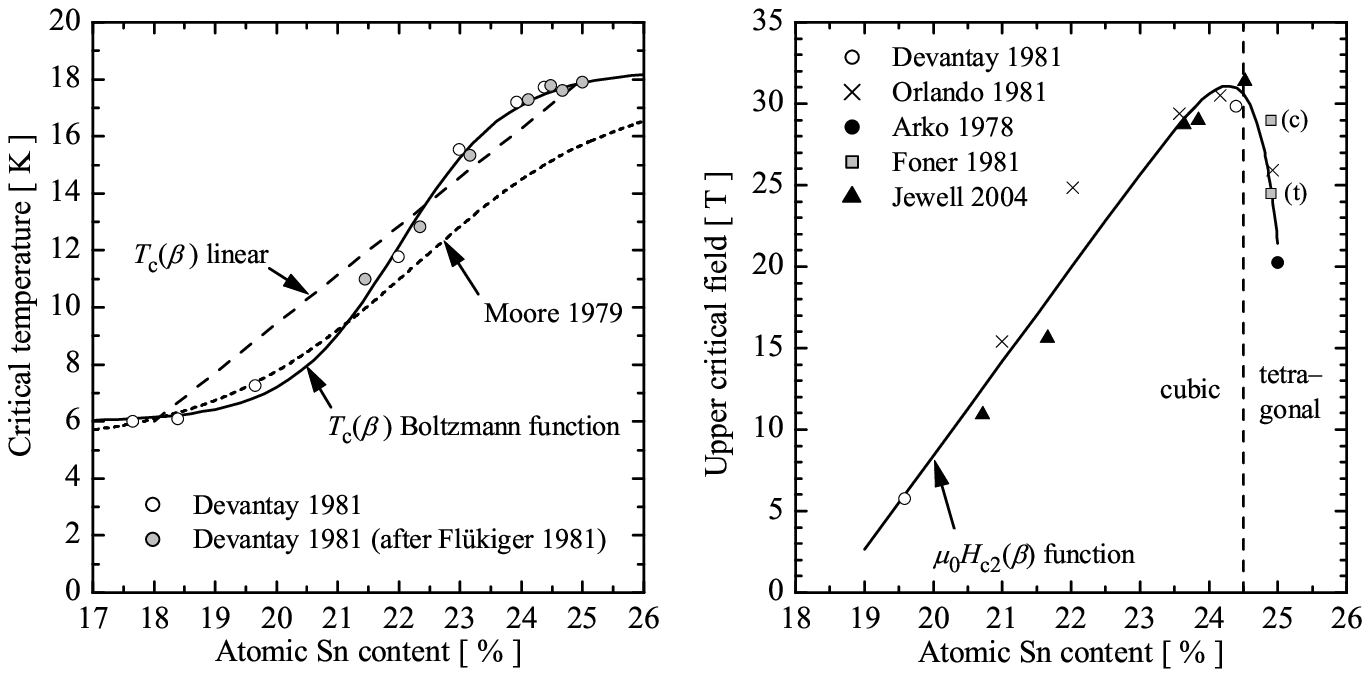}
\caption{\label{tcandhc2beta} Literature results for the critical temperature (left pot) and upper critical field at zero temperature (right plot) as function of Nb-Sn composition. The $\tc(\beta)$ Boltzmann function (\ref{boltzmanntc}) and the $\mhc(\beta)$ function (\ref{hc2beta}) are empirical relations that summarize the available literature results.\\\scriptsize{\textit{Adapted from~\cite{Devantay1981JMS} \copyright 1981 Chapman and Hall Ltd., \cite{Jewell2004ACEM} \copyright 2004 The American Institute of Physics, \cite{Flukiger1982BUCKELBOOK}, \cite{Moore1979PRB} \copyright 1979 The American Physical Society. With kind permission of Springer Science and Business Media, the American Physical Society, the American Institute of Physics, R.~Fl\"{u}kiger, T.P.~Orlando, M.R.~Beasley and M.C.~Jewell.}}}
\end{figure*}

The results from Devantay \etal are reproduced in the left plot in figure~\ref{tcandhc2beta}. Some points are reproduced from Fl\"{u}kiger~\cite{Flukiger1982BUCKELBOOK}, who credited these (partly) additional results also to Devantay \etal. The linear $\tc(\beta)$ fit, indicated by the dashed line, was originally proposed by Devantay \etal to summarize their results:
\begin{equation}\label{devantaytc}
    \tc \left( \beta  \right) = \frac{{12}}{{0.07}}\left( {\beta  - 0.18} \right) + 6.
\end{equation}
The dotted curve summarizes the results of Moore \etal using (\ref{boltzmann}). The general tendency of the data sets of Devantay \etal and Moore \etal is similar, although the latter covers a wider A15 stability range and its $\tc$ values are slightly lower. The solid curve represents a fit to the data of Devantay \etal according to a Boltzmann function identical to (\ref{boltzmann}):
\begin{equation}\label{boltzmanntc}
    \tc \left( \beta  \right) = \frac{{ - 12.3}}{{1 + \exp \left( {\frac{{\beta  - 0.22}}{{0.009}}} \right)}} + 18.3.
\end{equation}
Equation (\ref{boltzmanntc}) assumes a maximum $\tc$ of 18.3~K, the highest recorded value for Nb$_3$Sn~\cite{Hanak1964RCA}.

The right plot in figure~\ref{tcandhc2beta} is a reproduction of a $\hc(\beta)$ data collection that was made by Fl\"{u}kiger~\cite{Flukiger1982BUCKELBOOK} with some modifications. After Fl\"{u}kiger, it represents a collection of results from Devantay \etal~\cite{Devantay1981JMS}, Orlando \etal~\cite{Orlando1979PRB,Orlando1981TM} and a close to stoichiometric single crystal from Arko \etal~\cite{Arko1978PRL}. The dotted line in the plot separates the cubic and tetragonal phases at 24.5 at.\% Sn. The Foner and McNiff results in figure~\ref{tcandhc2beta} are, in contrast to Fl\"{u}kiger's collection, here separated in the cubic and the tetragonal phase with assumed identical composition. The composition that was used here was taken from the Fl\"{u}kiger collection, which showed a single $\mhc$ point at a value of about 27~T (the average between the cubic and tetragonal phase). It is unclear where this compositional value is attributed to or whether the cubic single crystal in fact has a lower Sn content.

In addition to the results collected by Fl\"{u}kiger also recent $\hc(\beta)$ measurements from Jewell \etal~\cite{Jewell2004ACEM} are added. A slightly different route was followed than given in \cite{Jewell2004ACEM} for the analysis of the $\hc(T)$ results. The compositions of the homogenized bulk samples in \cite{Jewell2004ACEM} were calculated from the $\tc$ values assuming a linear $\tc(\beta)$ dependence as given by (\ref{devantaytc}). The resulting $\hc(\beta)$ using these compositions is linear and deviates from the collection by Fl\"{u}kiger, as shown in \cite{Jewell2004ACEM}. Here, the compositions were recalculated based on the $\tc$ values using the Boltzmann fit on the Devantay \etal results as given by (\ref{boltzmanntc}). The Jewell \etal bulk results then become consistent with the earlier $\hc(\beta)$ data. The highest $\mhc$ value of 31.4~T represented, although extrapolated~\cite{Jewell2004ACEM}, a record for binary Nb-Sn, but has very recently been surpassed by an (extrapolated) $\mhc(0)=35.7~\rm{T}$ in highly disordered bulk samples~\cite{Cooley2006APL}. The existing $\hc(\beta)$ data can be summarized by a summation of an exponential and a linear fit according to:
\begin{equation}\label{hc2beta}
    \mu _0 \hc \left( \beta  \right) =  - 10^{ - 30} \exp \left( {\frac{\beta }{{0.00348}}} \right) + 577\beta  - 107.
\end{equation}
The solid curve in the right graph in figure~\ref{tcandhc2beta} is calculated using (\ref{hc2beta}). It is interesting to note that a linear extrapolation of the cubic $\hc(\beta)$ results yields around 35~T for the stoichiometric composition, in agreement with the reported record value by Cooley \etal~\cite{Cooley2006APL}. Unfortunately, however, compositional information is not available for this bulk sample, a high degree of uncertainty is stated for their reported resistivity value of 300~$\mu\Omega$cm, and the corresponding $\tc$ value of 15.1~K would place the Sn content around 23\%. Nevertheless, (\ref{boltzmanntc}) and (\ref{hc2beta}) summarize the consistent existing literature data on well defined, homogeneous laboratory samples for the main superconducting parameters $\tc$ and $\hc$ as function of composition.
%
%-----------------------------------------------------------------------------------------------------------
\section{\label{hc2tbeta} Changes in $\hc(T)$ with atomic tin content}
To describe the critical currents in wires, not only the upper critical field at zero temperature and zero field critical temperature are important. Also an accurate description of the entire field-temperature phase boundary is required. The temperature dependence of the upper critical field as function of temperature $[\hc(T)]$ is well investigated up to about $0.5\hc$, mainly since the field range up to about 15~T is readily available using standard laboratory magnets. Behavior at higher fields is often estimated by using an assumed  $\hc(T)$ dependence or by using so called 'Kramer'~\cite{Kramer1973JAP} extrapolations of lower field critical current data that rely on an assumed pinning mechanism. Estimates of $\mhc(0)$ or Kramer extrapolated critical fields $[\mu_0H_{\rm{K}}(0)]$ in wires that are derived in this way range from 18~T to $>$~31~T \cite{Filamentarybook1980,Kroeger1980JAP,Hechler1969JLTP,Suenaga1986JAP,Akihama1977TM,Tachikawa1981APL,Lindenhovius2000TAS,tenHaken1999JAP,Cheggour2002CRY,Godeke2003SUST}. It is obvious that this wide range of claimed $\hc(0)$ values in wires complicates understanding of their behavior. It is therefore required to analyze data over the full field range in better defined samples. Three data-sets which are measured over nearly the full field-temperature range on well defined laboratory specimen exist in the literature which will be outlined below. Recent investigations on $\hc(T)$ over the entire magnetic field range in wires emphasized on the compositional dependence and the shape of the phase boundary~\cite{Godeke2005JAP,Godeke2005PHD}. It was shown that all available $\hc(T)$ can be described by one single relation, indicating that the shape of the field-temperature phase boundary is constant and independent of composition, morphology or strain state. Knowledge of $\hc(0)$ and $\tc(0)$ thus means that the entire phase boundary is known. It was concluded that the highest Sn fractions in wires represent an upper phase boundary of $\mhc(0) \cong 29~\rm{T}$ and $\tc(0) \cong 18~\rm{K}$ for ternary wires. The critical current scales with a lower phase boundary which is an effective average over the compositional range that is present in a wire.
%
%----------------------------------------------------------------------------------
\subsection{\label{hc2tcubictetragonal} $\hc(T)$ of cubic and tetragonal phases in single- and polycrystalline samples}
\begin{figure*}
\includegraphics [scale=1]{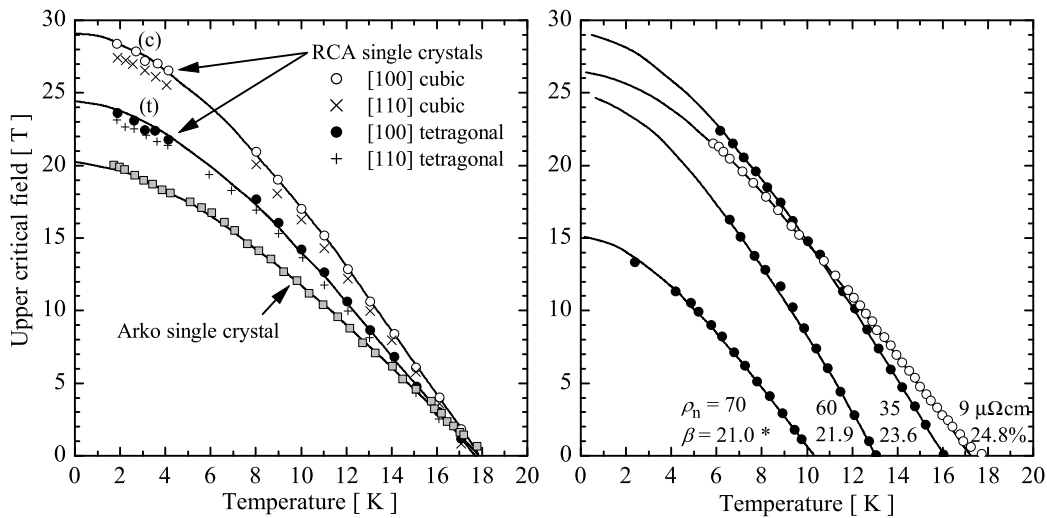}
\caption{\label{hc2tfonerorlando}  The upper critical field versus temperature for single crystals after Foner and McNiff~\cite{Foner1981SSC}$^{\dag}$ at approximately a 50 to 90\% resistive criterion (left plot), and for thin films with varying resistivity after Orlando \etal~\cite{Orlando1981TM}$^{\ddag}$ at a 50\% resistive criterion (right plot). *Compositions were calculated from the resistivity values using (\ref{rhobeta}).\\$^{\dag}$\scriptsize{\textit{\copyright 1981 Pergamon Press Ltd. There are instances where we have been unable to trace or contact the copyright holder or the original authors. If notified the publisher will be pleased to rectify any errors or omissions at the earliest opportunity.}}\\$^{\ddag}$\scriptsize{\textit{\copyright 1981 IEEE. Adapted with kind permission of IEEE and T.P.~Orlando.}}}
\end{figure*}

The articles by Foner and McNiff~\cite{Foner1976PL,Foner1981SSC} contain data over the entire magnetic field range on single- and polycrystals in the cubic and tetragonal phase. In addition, single crystals are investigated in the [100] and [110] directions to analyze the anisotropy in $\hc(T)$. These results are reproduced in the left plot in figure~\ref{hc2tfonerorlando}, together with the results of Foner and McNiff on the approximately stoichiometric Arko \etal single crystal~\cite{Arko1978PRL}. Their characterizations were performed using a RF technique and the resulting $\hc(T)$ are claimed to correspond to a 50 to 90\% resistive criterion. The curves are fits to the data (in the [100] directions) according to microscopic theory assuming a dirty Type II superconductor and no Pauli paramagnetic limiting. The anisotropy in $\hc(T)$ is 3 to 4\%, both in the cubic and in the tetragonal phase. The shape of the $\hc(T)$ phase boundary and $\tc(0)$ are similar despite the 9~T difference in $\mhc(0)$ between the three crystals. The $\mhc(0)$ value for the tetragonal phase is about 4.5~T lower than for the cubic phase. A slightly smaller difference in $\mhc(0)$ of about 3.5~T was observed by Foner and McNiff for polycrystalline sample material in the cubic and tetragonal phase. The reduced $\hc(0)$ of the (partly?) tetragonal polycrystalline sample was combined with a 0.1 to 0.2~K rise in $\tc(0)$. Their cubic polycrystalline sample averaged the [100] and [110] directions of the cubic single crystal. The single- and poly-crystal results of Foner and McNiff can explain differences in $\hc(0)$ values for Sn rich A15 compounds in terms of differences between the cubic and tetragonal phases.
%
%----------------------------------------------------------------------------------
\subsection{\label{hc2tthinfilms} $\hc(T)$ of thin films with varying resistivity}
Thin film $\hc(T)$ results on films with varying resistivity are available from Orlando \etal~\cite{Orlando1979PRB,Orlando1981TM} for fields up to 23~T. They used a resistive technique with a 50\% normal state criterion. Their results are reproduced in the right plot in figure~\ref{hc2tfonerorlando}. The curves are fits to the data according to microscopic theory in the absence of Pauli paramagnetic limitations. Orlando \etal quote the resistivities of the films. In figure~\ref{hc2tfonerorlando} also the compositions are given, which were calculated from the resistivity values using (\ref{rhobeta}). Lowering the resistivity, or increasing the Sn content, increases the field-temperature phase boundary. Low resistivity ($9~\mu\Omega\rm{cm}$), close to stoichiometric (24.8 at.\% Sn) A15 has a $\tc(0)$ of close to 18~K, but a reduced $\mhc(0)$ of about 26.5~T. An apparently optimal dirty thin film ($35~\mu\Omega\rm{cm}$, 23.6 at.\% Sn) reaches $\mhc(0)\cong 29~\rm{T}$ at the cost of a reduced $\tc(0)\cong16~\rm{K}$. The lowest phase transition that was measured occurred for $\rhn=70~\mu\Omega\rm{cm}$ ($\beta=21$ at.\% Sn) at $\mhc(0)\cong15~\rm{T}$ and $\tc(0)\cong10.5~\rm{K}$. The shape of the phase boundaries are again similar and independent of composition. The Orlando \etal results yield explanations for the placement of the phase boundary in terms of resistivity and composition.
%
%--------------------------------------------------------------------------------
\subsection{\label{hc2tbulk} $\hc(T)$ of bulk materials}
\begin{figure}
\includegraphics [scale=1]{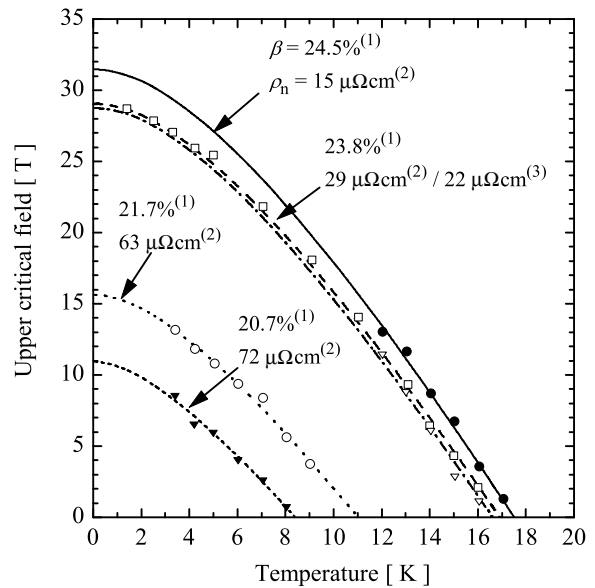}
\caption{\label{jewellhc2t} Upper critical field as function of temperature on homogenized bulk samples with varying composition after Jewell \etal~\cite{Jewell2004ACEM}$^{\dag}$ at approximately a 90\% resistive criterion. (1) Compositions calculated from $\tc(0)$ using (\ref{boltzmanntc}). (2) Resistivities calculated from compositions (1) using (\ref{rhobeta}). (3) Resistivity extrapolated from $\rhn(H)$ data.\\$^{\dag}$\scriptsize{\textit{\copyright 2004 The American Institute of Physics. Adapted with kind permission of the American Institute of Physics and M.C.~Jewell.}}}
\end{figure}

New $\hc(T)$ for the full field range has recently become available on bulk samples from Jewell \etal~\cite{Jewell2004ACEM}. Their results are reproduced in figure~\ref{jewellhc2t}. The curves are fits to the data according to microscopic theory using the Maki-de~Gennes approximation~\cite{Godeke2005JAP}. The depicted compositions and resistivities were calculated from the $\tc(0)$ values using (\ref{boltzmanntc}) and (\ref{rhobeta}). All samples were measured in a Vibrating Sample Magnetometer (VSM) using the first observable diamagnetic deviation from reversible normal state paramagnetic behavior as criterion for $\hc$ (which approximately corresponds to a 90\% resistive criterion~\cite{Godeke2005JAP}), except the $\beta=23.8$~at.\% Sn sample which was measured resistively, using a 90\% normal state criterion. All magnetically measured bulk samples were homogenized but not the resistively measured sample. Its measured resistivity value (extrapolated to zero applied magnetic field from $\rhn(H)$ data) amounted $22~\mu\Omega\rm{cm}$. That this value is lower than the calculated value of $29~\mu\Omega\rm{cm}$ (from $\tc$) can be attributed to inaccuracies in voltage tap separation length which were about 30\% and/or percolation of the measuring current. These results are in qualitative agreement with the thin film data of Orlando \etal in the previous section, i.e. the field-temperature phase boundary increases dramatically if the A15 Sn content is increased. The shift of the boundary with composition appears proportional, i.e. without the tilt that occurred in the Orlando \etal results between the low resistivity and optimal dirty thin film. Again, the shape of the phase boundary is independent of composition.
%
%
%------------------------------------------------------------------------------
\subsection{\label{Concludingremarks}Concluding remarks}
The three $\hc(T)$ data sets on defined laboratory samples are highly consistent. Bulk sample materials are arguably the closest representation of A15 layers in wires. More experiments on bulk materials will therefore be welcome to clarify the inconsistency between the $\tc(\beta)$ data sets of Moore \etal and Devantay \etal and to generate a larger quantity and more accurate $\hc(\beta)$ database using the these days readily available high field magnet systems. The existing data can be summarized by plotting their extremes [$\hc(0)$ and $\tc(0)$] as function of composition as was done in figure~\ref{tcandhc2beta}. This is valid since the shape of the phase boundary is constant and independent of composition (and thus independent of the electron phonon interaction strength), or whether the material is in the cubic or the tetragonal phase~\cite{Godeke2005JAP,Godeke2005PHD}. The entire field-temperature dependence of the Nb-Sn can thus be summarized to depend solely on composition using (\ref{boltzmanntc}) and (\ref{hc2beta}) as was recently postulated~\cite{Godeke2005JAP}.
%
%------------------------------------------------------------------------------------------------------
\section{\label{additions} Copper, Tantalum and Titanium additions to the A15}
\subsection{\label{copper} Copper additions}
Wires are different from pure binary Nb-Sn systems since they are always at least ternary due to the presence of Cu. The addition of Cu lowers the A15 formation temperature, thereby limiting grain growth and thus retaining a higher grain boundary density which is required for pinning (see section~\ref{grainsize}). Understanding of how Cu catalyzes the reaction requires detailed knowledge of ternary phase diagrams. Although ternary phase diagrams of the Nb-Sn-Cu system are sparsely available in the literature (mostly at fixed temperatures)~\cite{Hopkins1977MT,Zwicker1975ZM,Neijmeijer1987ZM,Lefranc1976JLCM,Ageeva1982DIA,Neijmeijer1988PHD}, they do not complement and in regions contradict each other. It can therefore be argued that the exact influence of Cu additions is not thoroughly known. It is clear that the presence of Cu destabilizes the line compounds NbSn$_2$ and Nb$_6$Sn$_5$~\cite{Fonerbook1981,Smathers1990MH}. It was further suggested that the presence of the Nb-Sn melt is extended to far below 930~$^\circ$C, enabling rapid A15 formation~\cite{Lefranc1976JLCM}. Very low temperature A15 formation has been recorded at temperatures down to about 450~$^\circ$C in a ternary system containing 5.4 at.\% Cu~\cite{Lefranc1976JLCM}. Although Cu can be detected in the A15 layers~\cite{Livingston1977PSSA,Smathers1982ACEM}, it is generally assumed to exist only at the grain boundaries and not to appear in the A15 grains~\cite{Suenaga1983APL,Suenaga2004UNP}. The absence of Cu in the A15 grains might explain why the binary A15 phase diagram can often be used to, at least qualitatively, interpret compositional analysis in wires. Also, to first order, the addition of Cu does not dramatically change the superconducting behavior of wires as compared to binary systems~\cite{Godeke2005JAP}. Finally, the presence of Cu was recently stated to be a possible origin for $\hc(0)$ suppression in wires~\cite{Jewell2004ACEM}.
%
%-----------------------------------------------------------------------------------------------------------
\subsection{\label{tiandta} Titanium and Tantalum additions}
Possibilities to improve the superconducting properties of the Nb-Sn system by the use of third and fourth element additions have been extensively investigated (see~\cite{Drost1985TM,Akihama1977TM,Tachikawa1981APL,Livingston1978TM,Springer1984ACEM,Suenaga1985TM,Tachikawa1982ACEM,DewHughes1977TM,Togano1979JLCM,Suenaga1974JAP,Takeuchi1981CRY,Sekine1983TM,Wu1983TM,Kurahashi2005TAS,Abacherli2005TAS}) and reviewed~\cite{Flukiger1987TECH,Suenaga1986JAP}. Of a wide range of possible ternary additions Ta and Ti are most widely applied in wires. Both Ta and Ti occupy Nb sites in the A15 lattice~\cite{Tafto1984JAP}. The effect of these third elements is to suppress the low temperature cubic to tetragonal transition for $\geq$~2.8 at.\% Ta or for $\geq$~1.3 at.\% Ti (see e.g.~\cite{Flukiger1987TECH} and the references therein). Increasing Ta or Ti content has shown to increase deviations from stoichiometry which in itself is sufficient for stabilization of the cubic phase (figure~\ref{tcandhc2beta}). Retaining the cubic phase results in an increased $\hc(0)$, while retaining $\tc(0)$ (\cite{Foner1981SSC}, figure~\ref{hc2tfonerorlando}). Also the resistivity increases with increasing Ta or Ti additions~\cite{Flukiger1987TECH} which is often referred to as the main cause for an increase in $\hc(0)$. The latter is, according to the GLAG theory in the dirty limit, proportional to $\rhn$ through (Tinkham~\cite{Tinkham1996BK}, p.~162, modified using $1/\nu_{\rm{F}}\ell = 1/3e^2\nef\rhn$):
\begin{equation}\label{hc2rho}
    \mu _0 H_{{\rm{c2}}} \left( 0 \right) \cong k_{\rm{B}} e N\left( {E_{\rm{F}} } \right)\rho _{\rm{n}} \tc \left( 0 \right) = \frac{{3e }}{{\pi ^2 k_{\rm{B}} }}\gamma \rho _{\rm{n}} \tc \left( 0 \right),
\end{equation}
where $\nu_{\rm{F}}$, $\ell$ and $e$ represent the velocity of the electrons at the Fermi level, the electron mean free path and the elementary charge quantum respectively and $\gamma$ is the electron specific heat constant or Sommerfeld constant. Orlando \etal~\cite{Orlando1981TM} showed for pure binary films, that $\hc(0)$ [and $\hc(4.2~\rm{K})$] both increase with resistivity, peak at $\rhn\cong 25~\mu\Omega\rm{cm}$ and then start to decrease. Similarly, $\hc(4.2~\rm{K})$, as measured in alloyed wires~\cite{Suenaga1986JAP} initially increases with Ti or Ta additions, peaks at about 1.5 at.\% Ti or 4 at.\% Ta, and then reduces with increasing alloy content. The $\tc(0)$ in the pure binary films is about 18~K in the clean limit and initially not influenced by increasing resistivity. However, $\tc(0)$ starts to decrease above $\rhn\cong 30~\mu\Omega\rm{cm}$. At  $\rhn= 35~\mu\Omega\rm{cm}$, $\mhc(0)$ is still high at 29~T, but $\tc(0)$ has lowered to 16~K (figure~\ref{hc2tfonerorlando}). The critical temperature in alloyed wires peaks only slightly at about 1 at.\% Ti and 2 at.\% Ta. A clear reduction in $\tc(0)$ occurs above about 1.5 at.\% Ti and 3.5 at.\% Ta.

The above summary shows that the effects of Ta and Ti additions are similar but Ti is approximately a factor two more effective in increasing the resistivity. Indeed 1.3 at.\% Ti addition results in $\rhn\cong 40~\mu\Omega\rm{cm}$ whereas 2.8 at.\% Ta addition results in $\rhn\cong 30~\mu\Omega\rm{cm}$~\cite{Flukiger1987TECH}. Consistent overall behavior emerges when the initial increase in  $\hc(0)$ [while retaining a constant $\tc(0)$] with alloying, can be attributed to a suppression of the tetragonal phase, and above $\rhn\cong 30~\mu\Omega\rm{cm}$ (or about 1.5 at.\% Ti and 3.5 at.\% Ta) the increased resistivity suppresses both $\hc(0)$ and $\tc(0)$. Note that this suggestion is contrary to the general assumptions in the literature that $\hc(0)$ rises as a result of increased resistivity due to its proportionality with $\rhn$. It can be argued that such an assumption represents an oversimplification, since it neglects the occurrence of the ternary phase and contradicts combination of the well established datasets of $\rhn(\beta)$, $\hc(\beta)$ and $\hc(T)$: Increasing the resistivity at compositions below 24.5 at.\% Sn strongly \textit{reduces} $\hc(0)$ as follows from right plot in figure~\ref{tcandhc2beta}, since increasing $\rhn$ is equivalent to reducing the Sn content (figure~\ref{rhoSncontent}). The latter can be physically understood in terms of a reduction of the d-band peaks and thus in $\nef$ in (\ref{hc2rho}), which influences $\hc(0)$ directly, as well as through $\lambda_{\rm{ep}}$ in $\tc(0)$ [see (\ref{bcstc})]. This can counteract the increase of $\hc(0)$ through $\rhn$. The binary summary (figure~\ref{tcandhc2beta}) remains consistent, at least qualitatively, with Ta or Ti ternary additions if it is assumed that their main effect is to push the composition increasingly off-stoichiometric and simultaneously increasing the resistivity and preventing the tetragonal transformation. It should be emphasized that this argument is not well supported, since all knowledge on ternary systems with Ta and Ti stems from inhomogeneous wires as opposed to homogenized laboratory samples.
%
%-----------------------------------------------------------------------------------------------------------
\section{\label{grainsize} Grain size and the maximum bulk pinning force}
\begin{figure}
\includegraphics [scale=1]{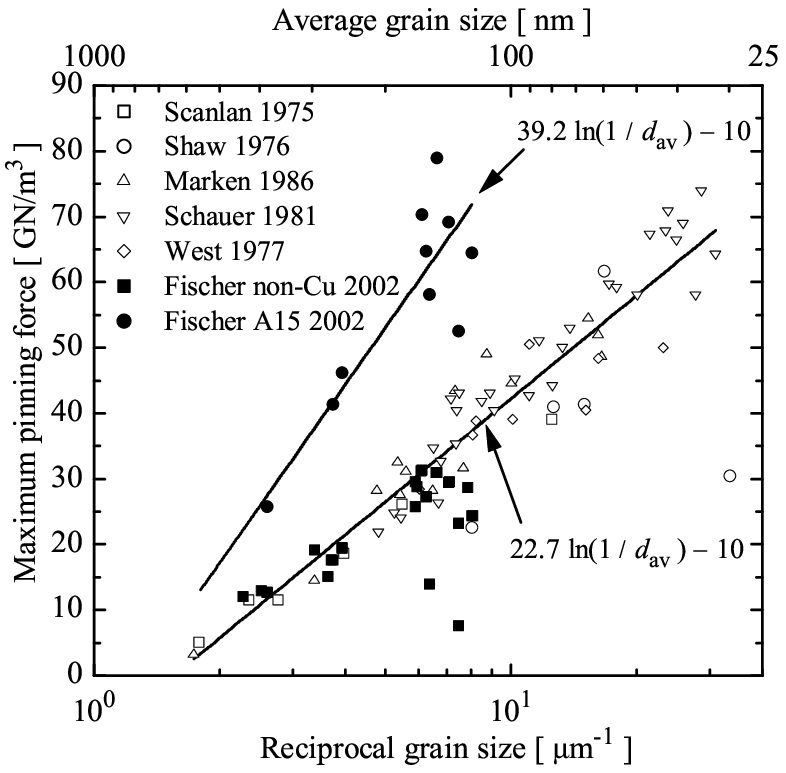}
\caption{\label{fpgrainsize} Maximum pinning force as function of reciprocal grain size after Fischer~\cite{Fischer2002MST}.\\\scriptsize{\textit{Adapted with kind permission of C.M.~Fischer.}}}
\end{figure}

A15 Nb-Sn behaves as a Type-II superconductor. Its current carrying capabilities for any practical field (above  $\mu_0H_{\rm{c1}}\cong 38~\rm{mT}$~\cite{Guritanu2004PRB}) thus rely on its capability to pin the flux-lines. The bulk pinning force at $\jc$ can be calculated from its critical balance with the Lorentz force:
\begin{equation}\label{pinningforce}
    {\bf{F}}_{\bf{P}}  \equiv  - {\bf{J}}_{\bf{c}}  \times {\bf{B}}.
\end{equation}
The bulk pinning force depends on the magnetic field and mostly has a maximum which is positioned roughly between 0.1 to $0.4\hc$~\cite{Cooley2002ACEM,Godeke2006SUST}. The magnetic field value at which the maximum occurs depends on the details of the flux-line to lattice interactions~\cite{Kramer1973JAP,Cooley2002ACEM,DewHughes1974PM}. From (\ref{pinningforce}) it is clear that the bulk pinning force determines the conductors' current carrying capabilities. The general assumption is that the grain boundaries in the A15 phase represent the primary pinning centers. This is supported by strong experimental evidence in which the maximum pinning force is related to the reciprocal grain size and thus the grain boundary density~\cite{Scanlan1975JAP,Shaw1976JAP,Marken1986PHD,Schauer1981TM,West1977JMS}. These results were summarized by Fischer~\cite{Fischer2002MST}, who also obtained additional results for the maximum bulk pinning force versus grain size from VSM characterizations. A reproduction after Fischer, including his own additions is shown in figure~\ref{fpgrainsize}. It remains unresolved how grain boundary pinning in Nb$_3$Sn occurs in detail. It is obvious that minima are present in the superconducting wave-function at the grain boundaries but it is unclear what physical principle (the presence of Cu, morphology, etc.) constitutes to this. Investigating what is present at the grain boundary might resolve some of this question and recent efforts are made in this respect~\cite{Flukiger2005MT17}.

The calculated value for $F_{\rm{P}}$ from transport current depends on the area for $\jc$, which can be normalized to e.g. the non-Cu area or the A15 area. Most pinning force data from the literature are calculated from the non-Cu $\jc$ and figure~\ref{fpgrainsize} shows that all pre-2002 results are consistent. They follow a single line on a semi-logarithmic plot summarized with:
\begin{equation}\label{fpmaxfit1}
    F_{{\rm{P, max}}}  = 22.7\ln \left( {\frac{1}{{d_{{\rm{av}}} }}} \right) - 10 ~~~ \left[ {{\rm{GN/m}}^{\rm{3}} } \right],
\end{equation}
in which $d_{\rm{av}}$ represents the average grain size in $\mu\rm{m}$. Additionally the maximum bulk pinning force data from Fischer, calculated from VSM derived non-Cu $\jc$ values are included in the graph (black squares). These results approximately correspond to the earlier literature data. If, however, the values obtained by Fischer are calculated from $\jc$ values which are normalized to the A15 area, the calculated maximum pinning force becomes higher due to the increased $\jc$ values. The slope in the semi-logarithmic plot effectively increases and Fischer's A15 area results can thus be summarized by:
\begin{equation}\label{fpmaxfit2}
    F_{{\rm{P, max}}}  = 39.2\ln \left( {\frac{1}{{d_{{\rm{av}}} }}} \right) - 10 ~~~ \left[ {{\rm{GN/m}}^{\rm{3}} } \right],
\end{equation}
in which again $d_{\rm{av}}$ is in $\mu\rm{m}$.

The most objective inter-wire comparisons can be made when the pinning force is normalized to the available grain boundary length resulting in an effective pinning force $Q_{\rm{GB}}$~\cite{Fischer2002MST}:
\begin{equation}\label{qgb}
    Q_{{\rm{GB}}}  = \left( {\frac{1}{\eta }} \right)\left( {\frac{{F_{\rm{P}} }}{{GB_{l/A} }}} \right),
\end{equation}
in which $\eta$ represents an efficiency normalization (assumed 1/3 for equiaxed grains) and $GB_{l/A}$ represents the grain boundary length per area. The effective pinning force can be interpreted as a measure of the effective pinning strength of a grain boundary. Calculation of (\ref{qgb}), however, requires an accurate determination of $GB_{l/A}$ which, although thoroughly analyzed by Fischer~\cite{Fischer2002MST}, is often not available.

The optimum grain size to achieve the ideal limit of one pinning site per flux-line can be estimated, since the grain boundaries are the primary pinning centers in Nb$_3$Sn. This limit will be approached if the grain size is comparable to the flux-line spacing at a given field. The flux-line spacing can be calculated by assuming a triangular flux-line lattice~\cite{Tinkham1996BK}:
\begin{equation}\label{fluxlinespacing}
    a_\vartriangle\left( H \right) = \left( {\frac{4}{3}} \right)^{{1 \mathord{\left/ {\vphantom {1 4}} \right.\kern-\nulldelimiterspace} 4}} \left( {\frac{{\phi _0 }}{{\mu _0 H}}} \right)^{{1 \mathord{\left/ {\vphantom {1 2}} \right. \kern-\nulldelimiterspace} 2}},
\end{equation}
which results in 14~nm at 12~T. The smallest grain sizes in figure~\ref{fpgrainsize} are about 35~nm but commercial wires have grain sizes of 100 to 200~nm and are thus far from optimized. In NbTi in contrast, the ideal limit of one flux-line per pinning site has been closely approached through specific heat treatments that introduce a fine distribution of normal conducting alpha-Ti precipitates that act as pinning sites~\cite{Lee2003WJI}. NbTi is therefore (under present understanding) close to fully optimized.
%
%-----------------------------------------------------------------------------------------------------------
\section{\label{strain} Variations of the superconducting properties with strain}
\subsection{\label{microscopicorigin} Microscopic origins of strain dependence}
The superconducting properties of all materials are to a certain extent sensitive to strain, although the effect can be marginal as for example for NbTi. The strain influences on the critical current in the Nb-Sn system originate from shifts in the field-temperature phase boundary, i.e. $\jc$ varies through a change in  $\hc(T)$ and possibly a direct influence on the maximum bulk pinning force, as is detailed elsewhere~\cite{Godeke2006SUST}. Strain influences can be separated into two regimes. Obviously, large strain can evoke serious and irreversible damage to the superconductor in the form of cracks in the brittle A15 material~\cite{Jewell2003SUST}. Lower strain can vary the superconducting properties through strain induced lattice instabilities that result in a tetragonal distortion~\cite{Savitskii1976PSSA,Flukiger1981TM} and by influencing the electron-electron interactions, which can both be reversible. The latter can, in the case of A15 Nb-Sn, be discussed in terms of lattice distortions that influence the Nb chain integrity and thus $\nef$, or in terms of changes in lattice vibration modes which influence the electron-phonon interaction spectrum [which includes changes in $\nef$].

\begin{figure}
\includegraphics [scale=1]{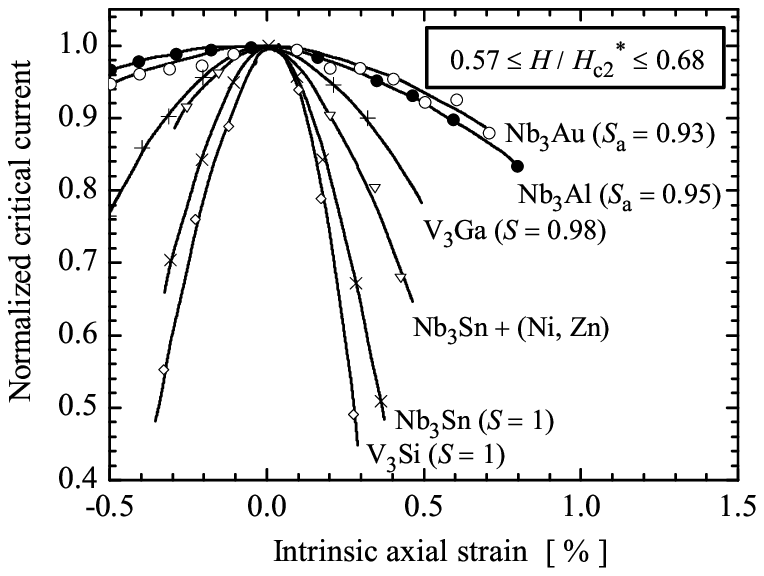}
\caption{\label{straindependence} Strain sensitivity of the critical current for different A15 superconductors with varying amounts of disorder after Fl\"{u}kiger \etal~\cite{Flukiger1984ACEM}$^{\dag}$.\\$^{\dag}$\scriptsize{\textit{\copyright 1984 Plenum Press. Adapted with kind permission of Springer Science and Business Media and R.~Fl\"{u}kiger.}}}
\end{figure}

Since distorting the lattice will modify $\nef$, it can be argued that the net effect is similar to an increase in the amount of disorder. This would render strain effects to be qualitatively similar to irradiation damage, or also similar to Sn deficiency if $\rhn(S)$ (figure~\ref{rhoorder}) and $\rhn(\beta)$ (figure~\ref{rhoSncontent}) can be considered of identical origin (section~\ref{Variationsinlatticeproperties}). Snead and Suenaga~\cite{Snead1980APL} have performed irradiation experiments on A15 filaments with and without pre-strain and concluded that irradiation increases the effect of pre-strain on $\tc$. This supports the above hypothesis, since the fact that pre-strain and irradiation amplify each other could imply that these effects are based on a similar microscopic principle [i.e. a modification of $\nef$]. Reversing this line of thought it can be argued that increasing the amount of disorder in a highly ordered system will have a relatively larger effect on its superconducting properties compared to a system which is already substantially disordered. Fl\"{u}kiger \etal collected strain sensitivity data of the critical current for various A15 wires with a varying degree of ordering and suggested to attribute changes in strain sensitivity to the degree of ordering~\cite{Flukiger1984ACEM}. His collection is reproduced in figure~\ref{straindependence}. Specifically, when comparing Nb$_3$Sn ($S=1$) and Nb$_3$Al ($S_{\rm{a}}=0.95$), the latter being always decidedly off-stoichiometric due to the instability of the stoichiometric phase at lower temperatures, the differences are large, despite roughly comparable $\hc$ and $\tc$ values. If the above argument is correct, it can be expected that strain sensitivity of the A15 Nb-Sn system depends on composition. Supporting this hypothesis is the fact that in axial strain experiments on wires, the strain sensitivity of the critical current increases with increasing magnetic field or temperature (see e.g.~\cite{Godeke2006SUST} and the references therein). At higher magnetic field and temperature, the lower Sn A15 fractions are no longer superconducting. With increasing magnetic field or temperature, the observed strain dependence is more and more determined by the higher Sn A15 fractions in the wires. These results therefore also represent a suggestion for enhanced strain sensitivity at higher Sn content. Whether this indeed is true is important for practical applications, since the present trend to improve the performance of commercial wires by Sn enrichment in the A15 could lead to an undesired increased strain sensitivity.

Strain sensitivity in terms of variations in electron-phonon interaction strength can be explained within the weak coupling BCS theory and also in the more general Eliashberg formalism by a pressure dependence of $\tc(0)$~\cite{TenHaken1994PHD}. However, strong indications exist throughout the literature that hydrostatic strain has only a negligible influence in comparison to non-hydrostatic or distortional strain components~\cite{TenHaken1994PHD,Testardi1971PRB,Welch1980ACEM}. Following these initial publications, which were based on strong coupling renormalized BCS theory, a recent attempt has been made to couple a microscopic full strain invariant analysis to Eliashberg-based relations for the critical temperature through strain-induced modifications in the electron-phonon spectrum~\cite{Markiewicz2004CRYa,Markiewicz2004CRYb,Markiewicz2005TAS}. This analysis is discussed in more detail elsewhere~\cite{Godeke2005PHD,Godeke2006SUST}. These recent descriptions are based on $\tc$ variation through a change in the electron-phonon interaction constant by a change in the lattice vibration modes. The latter is modeled by strain induced modifications to the frequency dependent electron-phonon interaction (in which $\nef$ is implicit) and phonon DOS. These models were verified using literature results for $\tc$ versus hydrostatic strain on a single crystal and $\tc$ versus axial strain on wires and tapes. It should be emphasized, however, that for a precise description of the strain influences also the strain dependence of $\hc(T)$ should be correlated to the microscopic theory.

Strain induced changes in $\nef$ in the full invariant analysis for $\tc$ are only accounted for through a strain-modified frequency dependence of the electron-phonon interaction, i.e. the change in $\nef$ is not directly calculated, but this apparently yields sufficient accuracy. For a microscopic description of the strain dependence of $\hc(T)$, however, the exact change in $\nef$ with strain should be known, as is evident from (\ref{hc2rho}). This direct influence of a change in $\nef$ on $\hc(0)$ would hypothetically result in a higher strain sensitivity for $\hc(0)$ compared to $\tc(0)$.  This is verified by experimental results where the relative strain dependence of $\hc(0)$ is roughly a factor 3 times higher than for $\tc(0)$ (see e.g.~\cite{Godeke2006SUST} and the references therein). This enhanced strain sensitivity of $\hc(0)$ compared to $\tc(0)$ was not satisfactory explained using strong coupling corrected BCS theory~\cite{Welch1980ACEM}. Promising recent work by Oh and Kim~\cite{Oh2006JAP}, that utilizes the interaction strength independent Eliashberg theory, yields a more satisfactory explanation, although preferably this should be combined with Markiewicz's his full invariant analysis to arrive at a complete description for the strain dependence of the superconducting properties of Nb$_3$Sn.
%
%-------------------------------------------------------------------------------------------------
\subsection{\label{straininsamples} Available strain experiments on well defined samples}
The previous sections showed that the main superconducting properties ($\tc$ and $\hc$) are well investigated on laboratory samples with defined composition. The available literature presents a clear and consistent composition dependence which differs only in details between the various results. Strain sensitivity experiments on defined laboratory samples are, in contrast to this, seriously lacking. Abundant results are available on technical wires which focus on the influences of axial strain on the critical current density (see e.g. the pioneering work by Ekin~\cite{Ekin1984ACEM} and the references in~\cite{Godeke2006SUST}). Wires are, however, due to their fabrication processes, always decidedly inhomogeneous in composition as well as strain. The foregoing discussions on the possible microscopic origins of strain dependence emphasized the need for strain dependence results, obtained from samples with a fixed and known composition, since both composition and strain influence $\hc(T)$. Results on laboratory specimen with defined composition are limited to hydrostatic experiments on single crystals or polycrystalline layers (e.g.~\cite{Chu1974JLTP,Lim1983PRB}) and one uniaxial experiment, also on a single crystal~\cite{McEvoy1971PHY}. These result in ${\rm{d}}\tc/{\rm{d}}P=-0.14$ and  $-0.22$~K/GPa for a single crystal~\cite{Chu1974JLTP} and a polycrystalline layer~\cite{Lim1983PRB} respectively. One result exists for the pressure dependence of $\hc$ on an inhomogeneous tape conductor~\cite{TenHaken1994PHD}, which resulted in ${\rm{d}}\mhc/{\rm{d}}P=-1.2\pm0.2$~T/GPa.

Next to inhomogeneities, also the three dimensionality of strain complicates the analysis in wires. Ten Haken~\cite{TenHaken1994PHD} made attempts to overcome this problem by switching to deformation experiments on two dimensional tape conductors. Although this has resulted in many new experimental insights that emphasize the importance of non-hydrostatic strain components, the tape conductors used in his research still suffered from compositional imperfections.

The lack of extensive strain experiments on well defined laboratory samples thus limits the ability to develop a description for $\tc$ and $\hc$ [or more specifically $\hc(T)$] versus strain, due to the possible different strain sensitivity with composition. Such results are, however, required to accurately analyze wire behavior. The data sets needed for this would preferably be analyzed in simplified model samples for which the three dimensional strain can be calculated and which separate the strain dependencies for well defined compositions.
%
%------------------------------------------------------------------------------------------------
\section{\label{conclusions} Conclusions}
Superconductivity in the A15 Nb-Sn intermetallic is microscopically well described in general terms of electron-phonon coupling of Cooper pairs. Strong coupling corrections to the microscopic theory become necessary above 23 at.\% Sn. Below this concentration the material is weak coupling. The A15 lattice contains Nb chains in which the atoms are closely spaced resulting in an increased density of states around the Fermi level, which is believed to be the main origin for the high critical temperature. The binary phase diagram is well established although a preference for the formation of close to stoichiometric A15 can be argued. Effects of off-stoichiometry can be discussed in terms of lattice parameter, long range ordering, resistivity or atomic Sn content. These terms are equivalent and consistent behavior is observed throughout the literature. The field-temperature phase boundary varies substantially with Sn content, but the observations can be explained in terms of off-stoichiometry and the occurrence of the tetragonal phase close to stoichiometry. The dependence of the critical temperature and upper critical field on compositional variations varies only in detail throughout the literature and is well established for the binary system. This consistent behavior arguably also remains valid when ternary additions such as Cu, Ti or Ta are present.

Consistent ternary or quaternary phase diagrams are not available but the pure binary phase diagram can be used, at least qualitatively for the A15 phase, to describe compositional variations in higher order systems. The addition of Cu allows for lower temperature A15 formation. It is present at A15 grain boundaries but does not appear to exist in the A15 grains. Additions of Ti and Ta occupy Nb sites and apparently result in distortions of the Nb chain integrity, cause off-stoichiometry, increase the resistivity and prevent the low temperature tetragonal transformation.

Abundant and consistent data on the bulk pinning force as function of reciprocal average grain size represent solid experimental evidence that the grain boundaries are the main pinning centers. The exact mechanism that determines the pinning strength at the grain boundaries remains unresolved. A consistent relation emerges between the maximum bulk pinning force, calculated from the critical current density normalized to the non-Cu area, and the average grain size. This maximum bulk pinning force, combined with the field-temperature phase boundary and strain effects, determines the attainable critical current density.

The influence of hydrostatic deformations can be qualitatively predicted from microscopic theory and limited data from hydrostatic deformation experiments are available. Distortional (non-hydrostatic) deformations, however, dominate the strain sensitivity. Although deformation experiments on wires are abundant, no general statements can be made. This is due to the lack of non-hydrostatic deformation experiments on homogeneous, and spatially simplified samples since it is plausible that strain sensitivity varies with composition. Very recent work opens new possibilities towards a full, strong coupling theory based description of strain sensitivity of the superconducting properties in the Nb-Sn system.
%
%------------------------------------------------------------------------------------------------------------
\section*{Acknowledgments}
Most of the material in this article was collected for the introduction chapters of my PhD thesis during my stay at the Applied Superconductivity Center, University of Wisconsin-Madison, USA and at the Low Temperature Division, the Special Chair for Industrial Application of Superconductivity at the Faculty of Science and Technology, University of Twente, Enschede, The Netherlands. I would like to thank Bennie ten Haken, Herman ten Kate and David Larbalestier for their rigorous corrections to the original text. This work was supported by the Netherlands Organization for Scientific Research (NWO) through the Technology Foundation STW and by the Director, Office of Science, High Energy Physics, U.S. Department of Energy under Contracts DMR--9632427 and DE--AC02--05CH11231. All contributions are greatly acknowledged.
%
%------------------------------------------------------------------------------------------------------------
%\section*{References}
%
%
\bibliographystyle{phaip}
\bibliography{Godeke}
\end{document}